# Enhanced water affinity of histidine by transition metal ions

Yongshun Song[1*], Jing Zhan[2*], Minyue Li[3], Hongwei Zhao[4,5], Guosheng Shi[3], Minghong Wu[3], Haiping Fang[1,4,5]

[1]*Department of Physics, East China University of Science and Technology, Shanghai 200237, China*

[2]*Department of Biomedical Engineering, Faculty of Engineering, National University of Singapore 117583, Singapore*

[3]*Shanghai Applied Radiation Institute, Shanghai University, Shanghai 200444, China*

[4]*Zhangjiang Laboratory, Shanghai Advanced Research Institute, Chinese Academy of Sciences, Shanghai 201210, China*

[5]*Shanghai Institute of Applied Physics, Chinese Academy of Sciences, Shanghai 201800, China*

Transitional metal ions widely exist in biological environments and are crucial to many life-sustaining physiological processes. Recently, transition metal ion such as $Cu^{2+}$, $Zn^{2+}$, $Ni^{2+}$, have been shown can increase the solubilities of aromatic biomolecules. Comparing with $Cu^{2+}$, $Zn^{2+}$ shows less enhancement to the solubilities of biomolecules such as tryptophan (Trp). On the other hand, $Zn^{2+}$ has a higher concentration in human blood plasma and appears in protein the most among transition metal ions, clarifying whether $Zn^{2+}$ can enhance the solubilities of other aromatic amino acids is significantly important. Herein, we observed that the solubility of aromatic amino acid histidine (His) is greatly enhanced in $ZnCl_2$ solution. Based on first principle calculations, this enhancement of solubility is attributed to cation-π interaction between His and $Zn^{2+}$. Our results here are of great importance for the bioavailability of aromatic drugs and provide new insights for the understanding of physiological functions of $Zn^{2+}$.

Most biochemical processes occur in the presence of water. Dispersion behavior of biomolecules in water has significance importance to their participation in physical, chemical and biological processes [1-4]. Aromatic moieties widely exists in biomolecules and drugs, which have been shown have a detrimental effect on aqueous solubility [5, 6]. As the simplest biomolecules and essential building blocks of proteins, the solubilities or hydrophilicity of aromatic amino acids are critical for protein folding, structure stabilization, and protein-ligand interactions [7-9]. Recently, we found that the cation-π interaction between transition metals and aromatic amino acids can greatly enhance the water affinity of aromatic amino acids, providing a new understanding for how metals modulate protein folding and influence protein energetics [10]. However, although some transition metal ions have the ability to enhance solubility, the range of enhancement varies from less than 2 times to more than 5 times under the same

---

mhwu@mail.shu.edu.cn; fanghaiping@sinap.ac.cn

[*] These authors contributed equally to this work.

concentration, according to the order of $Cu^{2+} > Ni^{2+} > Zn^{2+}$ (Figure 3 in Ref. [10]). On the other hand, based on a statistics of Protein Data Bank (PDB), there are about 8 times as many the number of proteins that bound with $Zn^{2+}$ as that bound with $Cu^{2+}$ [11]. $Zn^{2+}$ can involve in both structural and functional proteins [12]. Clarifying whether $Zn^{2+}$ can enhance the solubilities of other aromatic amino acids through cation-π interaction is significantly important.

Among all the aromatic amino acids (Tyr, Trp, Phe, His), His that contains an imidazole ring side chain is the one commonly involve in structure and function of biomolecules, since imidazole can easily participate in π-π interactions [13], hydrogen bond interaction [14], coordinate bond interaction [15] and cation-π interaction [16]. Besides, the amino acid His can also play an important role in biology, for example, $Cu^{2+}$-His complex has been found in human blood and is essential for $Cu^{2+}$ transport [17].

In this Letter, we found the cation-π interaction between $Zn^{2+}$ and His is significantly strong, showing a totally different behavior compared to interaction with Trp. By first principle calculations, we found that $Zn^{2+}$ can significantly enhance the water affinity of His, which is close to that of $Cu^{2+}$. Consistent results were obtained by solubility experiments, which show that the solubility of His can reach more than 5 times of that in pure water, in both 0.5 M $ZnCl_2$ and $CuCl_2$ solution. Our results here show that the strong cation-π interaction between His and $Zn^{2+}$ greatly affected the water affinity of of His, which will have further impact on the role of His played in protein folding, enzyme catalysis and structural stabilization of macromolecules.

To study the effect of $Zn^{2+}$ on the water affinity of His, we first employ the DFT computing technique, to calculate the interaction energy between the aromatic ring sidechain of His with (and without) $Zn^{2+}$ adsorption and the nearest neighboring water. This is a direct method to characterize the degree of polarity of aromatic sidechain, and has been used in previous work [10]. As shown in **Figure 1**(a), the interaction between sidechain of His and the nearest water is about -12.5 kcal/mol. For comparison, we also calculated the interaction between sidechain of Trp and the nearest water, which is about -9.5 kcal/mol, much smaller than that of His. The optimized distance between His (Trp) and the nearest water is 2.0 Å (2.2 Å) respectively, also showing that the interaction between His and the nearest water is stronger.

Water affinity is well related to solubility. Weak interactions with water means that the hydration of the compound is unfavorable, which contributes to their poor water solubility. Based on our previous computation, we expect that the solubilities of His and Trp in $ZnCl_2$ solution will show different manner. We then performed experiments on the solubility of His ($S_{His}$) in $ZnCl_2$ aqueous solution, and compared it with the solubility of Trp ($S_{Trp}$). The solubility of His in the presence of $ZnCl_2$ of different concentrations are measured. In **Figure 1**(b), it can be found that the solubilities of Trp and His in $ZnCl_2$ solution increase with a different manner (red rectangles in **Figure 1**(b)). For Trp, its solubility in $ZnCl_2$ solution increases by no more than 2 times, whereas for His, the solubility can increase to about 6 times as that in pure water. For comparison, we also measured the solubilities of His and Trp in CuCl$_2$ solution. With no surprise, Cu$^{2+}$ can both increase the solubilities of His and Trp over 5 times with a concentration of around 0.5 M. This result is consistent with our first principle computations on the adsorption energies (blue rectangles in **Figure 1**(b)), which show that the interaction between Zn$^{2+}$ and His is much greater than the interaction of Zn$^{2+}$ with Trp.

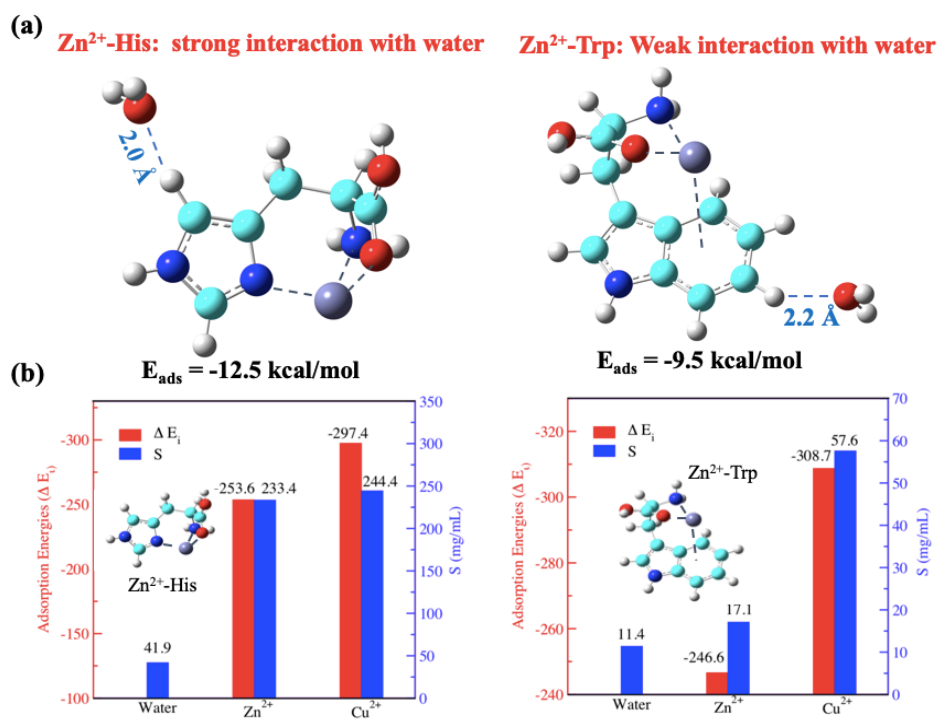

**Figure 1** (a) The optimized distance between Trp and the nearest water after Zn$^{2+}$ absorption is 2.2 Å and the interaction is -9.5 kcal/mol. In contrast, the optimized distance between His and the nearest water after Zn$^{2+}$ absorption is reduced to 2.0 Å and the interaction increases to -12.5 kcal/mol. (b) Left: Adsorption energies ($\Delta E_i$, red rectangles) of Cu$^{2+}$ and Zn$^{2+}$ with Trp, and water solubilities of Trp ($S_{Trp}$, blue rectangles) in 0.5 M ZnCl$_2$ and CuCl$_2$ aqueous solutions, respectively. (Reproduced from Ref. [10] with permission). Right: Adsorption energies ($\Delta E_i$, red rectangles) of Cu$^{2+}$ and Zn$^{2+}$ with His, and water solubility of His ($S_{His}$, blue rectangles) in 0.4 M ZnCl$_2$ and 0.4 M CuCl$_2$ aqueous solutions, respectively.

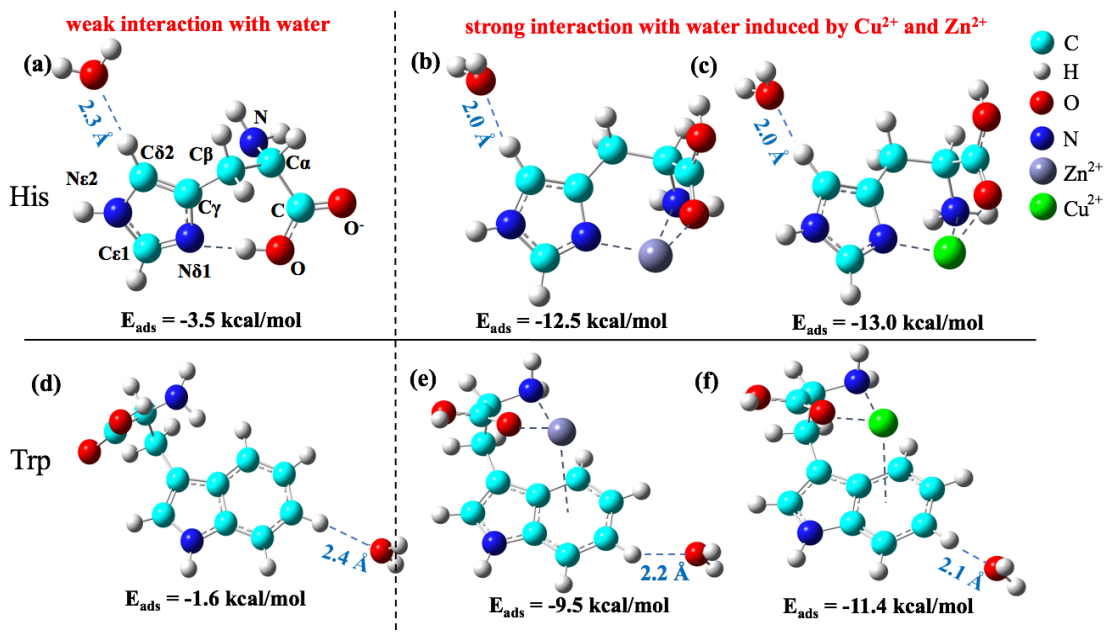

**Figure 2** The structure of His and Trp binding to $Zn^{2+}$ and $Cu^{2+}$ and the effect to the interaction between aromatic ring and the nearest water. (a) Optimized distance between imidazole and the nearest water is 2.3 Å, and the interaction between them is -3.5 kcal/mol. (b) The optimized distance between imidazole and the nearest water after $Zn^{2+}$ absorption reduced to 2.0 Å and the interaction increase to -12.5 kcal/mol. (c) The optimized distance between imidazole and the nearest water after $Cu^{2+}$ absorption reduced to 2.0 Å and the interaction increase to -13.0 kcal/mol. (d) Optimized distance between indole ring of Trp and the nearest water is 2.4 Å, and the interaction between them is -1.6 kcal/mol. (e) The optimized distance between indole and the nearest water after $Zn^{2+}$ absorption reduced to 2.2 Å and the interaction increase to -9.5 kcal/mol. (f) The optimized distance between indole and the nearest water after $Cu^{2+}$ absorption reduced to 2.1 Å and the interaction increase to -11.4 kcal/mol.

In $CuCl_2$ solution, both Trp and His show an obvious solubility increase, while they show a different manner in $ZnCl_2$ solution. This can be explained by our first principle calculations. As shown in **Figure 2**(a), originally the interaction between sidechain of His and the nearest water is about -3.5 kcal/mol. Whereas, when $Zn^{2+}$ is bound, it increases to about -12.5 kcal/mol (**Figure 2**(b)), about two times as large as hydrogen bond, very close to the strength of corresponding interaction when $Cu^{2+}$ is binding (-13.0 kcal/mol, as shown in **Figure 2**(c)). For comparison, the interaction between the indole ring of Trp and the nearest water is shown in **Figure 2**(d), which is about -1.6 kcal/mol. The interaction between indole ring and the nearest water is smaller than the corresponding interaction for His is consistent with the fact that Trp has a smaller solubility than His. **Figure 2**(e) and **Figure 2**(f) show the interaction between indole ring and the nearest water induced by $Zn^{2+}$ and $Cu^{2+}$ respectively, which both increase a large extent in comparison with the case of pure water. However, unlike His, the extent of interaction increased by $Zn^{2+}$ is not as large as by $Cu^{2+}$ (9.5 kcal/mol vs. 11.4 kcal/mol).

The optimized geometry of the metal-aromatic-water system calculated here is chosen based on that the nearest water forms hydrogen bond with aromatic molecule through π cloud. In fact, water molecules can interact in three possible ways: (1) an ion-dipole interaction, wherein the water interacts directly with the cation; (2) a π hydrogen-bond between water and the aromatic π cloud; (3) a NH-OH σ hydrogen bond. We do not consider the case (1), since it does not reflect the water affinity of aromatic molecule. When the metal ion interacting with aromatic molecules by cation-π interaction, the hydrogen bond (3) will be weakened, and the hydrogen bond (2) will be enhanced [18]. Because the water affinity of the whole aromatic molecules depends mainly on the most hydrophobic region, only the hydrogen bond (2) is considered here.

The solubility increase behavior respect to the concentration of $ZnCl_2$ or $CuCl_2$ solution was then investigated. As shown in **Figure 3**, the increase behavior of solubility can be fitted by, $S_A = A_M C_M + S_A^0$, where $A_M$ and $C_M$ are the water affinity factor of amino acid induced by metal ion $M^{2+}$ and the concentration of $M^{2+}$, $S_A$ and $S_A^0$ are the solubility of amino acid A in salt solution and pure water, respectively. The $A_M$ has a distinct physical meaning, which is the solubility of amino acid will increase by $A_M$ M for every mole salt concentration increase. For His, $A_{Cu}$ =4.86 and $A_{Zn}$=2.80, both are much greater than the corresponding solubility increase of Trp ($A_{Cu}$=0.46) [10].

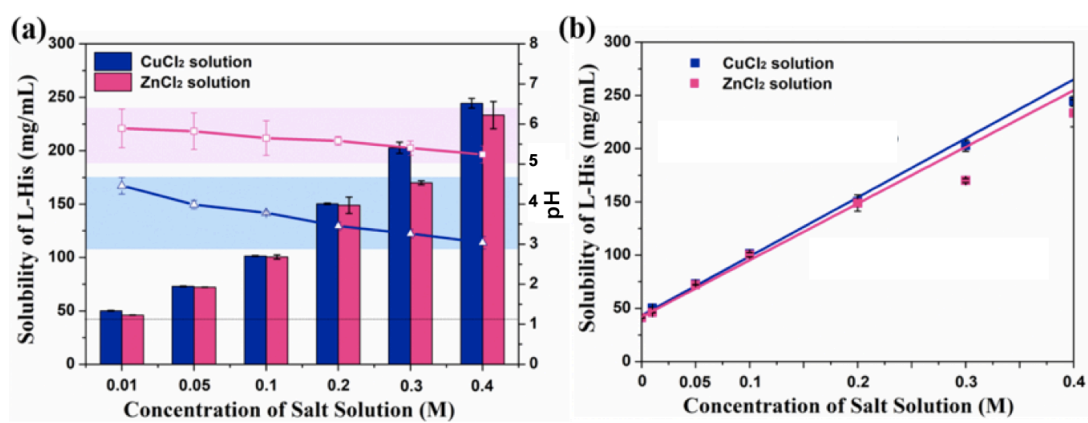

**Figure 3** (a) The solubility of L-His is increased gradually with the concentration of $CuCl_2$ or $ZnCl_2$, shown in the histogram. The concentration of $CuCl_2$ or $ZnCl_2$ and corresponding pH value are shown with red line and blue line. (b) The solubility of L-His increases monotonically with the concentration of $CuCl_2$ and $ZnCl_2$.

To valid the solubility increase mainly comes from the cation-π interaction between transition metal ions and the aromatic ring of Trp/His, the solubility of a non-aromatic amino acid glycine (Gly) was measured with the same condition. In **Figure 4**, it can be found that the solubilities of Trp and His both increase dramatically in $CuCl_2$ solution, while the solubility of Gly only shows a limited increase in the $CuCl_2$ solution. This result strongly support the idea that the higher solubility of the corresponding amino acid is mainly attributed to the aromatic ring of Trp/His that interacts with $Cu^{2+}$.

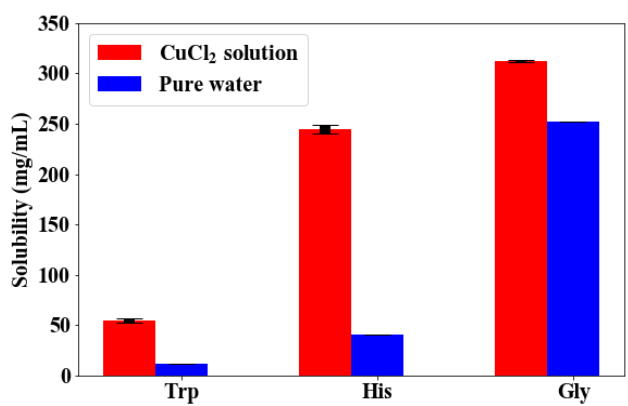

**Figure 4** Solubility of Trp, His and Gly in 0.4 M $CuCl_2$ aqueous solution (red rectangles) and pure water (blue rectangles).

The infrared (IR) spectra experiments on the powders (see Fig. S1) show structures of His in precipitates with $Cu^{2+}$ and $Zn^{2+}$ are the same with precipitates that without $Cu^{2+}$ or $Zn^{2+}$. Our results here is consistent with previous IR experiments for His [19].

NMR experiments on C12 in **Figure 5**(a) show that, $Cu^{2+}$-His have a clear chemical shift at 130 ppm region, which refer to the imidazole region of His [20]. $Zn^{2+}$-His also show a chemical shift at 130 ppm region, though with a much weaker signal (data not shown).

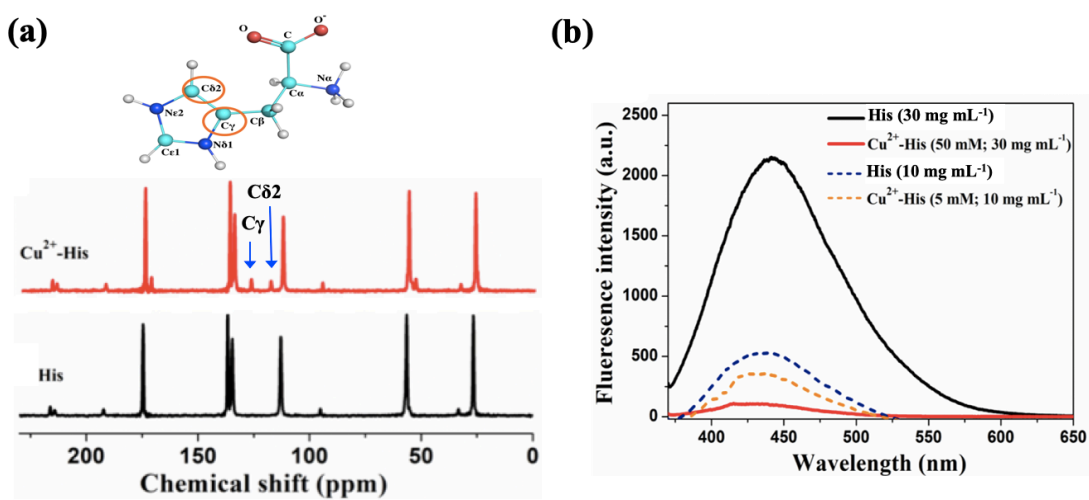

**Figure 5** (a) CP/MAS NMR C12 spectra of $Cu^{2+}$-His, and the control group (His in pure water). $C\delta2$ and $C\gamma$ peaks are characterized. (b) Fluorescence spectra of the His (10 mg ml$^{-1}$ and 30 mg mL$^{-1}$), and the His (30/10 mg mL$^{-1}$) in $CuCl_2$ solution (50/5 mM).

Fluorescence and UV absorption spectral experiments were performed to show evidence of the cation-π interaction between $Cu^{2+}$ and the imidazole ring of His. In **Figure 5**(b), the fluorescence spectrum of His excited at 360 nm has an emission peak at 448 nm, which is assigned to a conjugate double bond of the imidazole group that can easily generated the π-π* transition [21]. Compared with the fluorescence intensity of His in water, the intensity of His in the 50 mM $CuCl_2$ solution markedly decreased, indicating that the conjugate double bonds of the imidazole group in His are greatly affected in the $CuCl_2$ solution. Dashed line in **Figure 5**(b) showed that, this tendency is not affected when the concentration of His and $CuCl_2$ solution are decreased, only in a smaller range. These results demonstrate that the main effect of the fluorescence intensity of the conjugate double bonds of the imidazole group in His arises from $Cu^{2+}$ in $CuCl_2$ solution. Fluorescence spectrum of His in $ZnCl_2$ solution is shown in Fig. S2. Again, the fluorescence intensity of His and $ZnCl_2$ is quenched for $Zn^{2+}$-His complex.

The His contains six π electrons: four from two double bonds and two from a nitrogen lone pair. It can form π-π stacking interactions [13], though is complicated by the positive charge. It does not absorb at 280 nm in either cation state or neutral state, but does in the lower UV range (220 nm) more than some amino acids. Here, we observed that the UV absorption spectrum of His was also affected by the cation-π interactions between the imidazole ring of His and $Cu^{2+}$ in solution (shown in Fig. S3, which is an important evidence for the existence of cation-π interactions). Altogether, the fluorescence and UV absorption spectral experiments show the existence of cation-π interactions between the aromatic ring structure of His and $Cu^{2+}$ in solution, which is consistent with our theoretical prediction.

The enhancement of $S_{His}$ in $CuCl_2$ and $ZnCl_2$ solution does not come from the pH effect induced by hydrolysis of $Cu^{2+}$ or $Zn^{2+}$. As shown in **Figure 3**(a), the pH of $CuCl_2$ and $ZnCl_2$ solution both have a smaller decrease (for $Cu^{2+}$, from 4.5 to 3.1; for $Zn^{2+}$, from 6.0 to 5.2), indicating that pH change does not contribute the increased solubility of His.

With pH varies, there are different species exist for $Cu^{2+}$-His or $Zn^{2+}$-His complexes. The conformation we used for DFT computation is referred as $M(HL)^{2+}$ species in Ref. [22, 23]. This proposed tridentate coordination mode is the most accepted [24, 25].

We have previous showed that $Cu^{2+}$ ion binding to indole ring of Trp can increase the solubility of Trp [10]. This result is based on the novel experimental methods to obtain a high $Cu^{2+}$ concentration around Trp, otherwise $Cu^{2+}$ will form precipitates with Trp. We noted that His can easily bind to transition metal at Nδ1 site, independent of the experimental procedures. Though many studies on the complex of transition metal ion and His have been carried out [17], solubility change of His has seldom been studied.

In summary, our first principle calculations show that the cation-π interaction between $Zn^{2+}$ and His is strong and in $ZnCl_2$ solution the solubility of His can reach over 5 times that of His in pure water, very similar to the behavior of His in $CuCl_2$ solution, which is quite different from the cases with Trp previous reported. Theoretical studies show that the key to this unexpectedly experimental phenomenon is the strong cation-π interaction between the cations and the aromatic ring, which modifies the electronic distribution of the aromatic ring and enhances significantly the water affinity of the amino acids. Our results highlight the solubility enhancement of many imidazole derivatives by $Zn^{2+}$ may be a general phenomenon and need to draw more attention from research branches such as drug design, and artificial organic macromolecule design. The findings enrich the view of biomolecular solubility in aqueous electrolyte solution and provide new insights for the understanding of physiological functions of multivalent metal ions.

Supporting Information for

# Enhanced water affinity of histidine by transition metal ions

Yongshun Song[1*], Jing Zhan[2*], Minyue Li[3], Hongwei Zhao[4,5], Guosheng Shi[3], Minghong Wu[3], Haiping Fang[1,4,5]

[1]*Department of Physics, East China University of Science and Technology, Shanghai 200237, China*

[2]*Department of Biomedical Engineering, Faculty of Engineering, National University of Singapore 117583, Singapore*

[3]*Shanghai Applied Radiation Institute, Shanghai University, Shanghai 200444, China*

[4]*Zhangjiang Laboratory, Shanghai Advanced Research Institute, Chinese Academy of Sciences, Shanghai 201210, China*

[5]*Shanghai Institute of Applied Physics, Chinese Academy of Sciences, Shanghai 201800, China*

## Context

**PS1:** The spectra of the IR of $Cu^{2+}$–His, $Zn^{2+}$-His and His

**PS2:** Fluorescence spectra of the His, $ZnCl_2$, and $Zn^{2+}$-His

**PS3:** UV absorption spectra of $Cu^{2+}$–Trp solution

**PS4:** Experimental section

**PS5:** Computational methods

mhwu@mail.shu.edu.cn; fanghaiping@sinap.ac.cn

[*] These authors contributed equally to this work.

**PS1:** The spectra of the IR of $Cu^{2+}$–His, $Zn^{2+}$-His and His

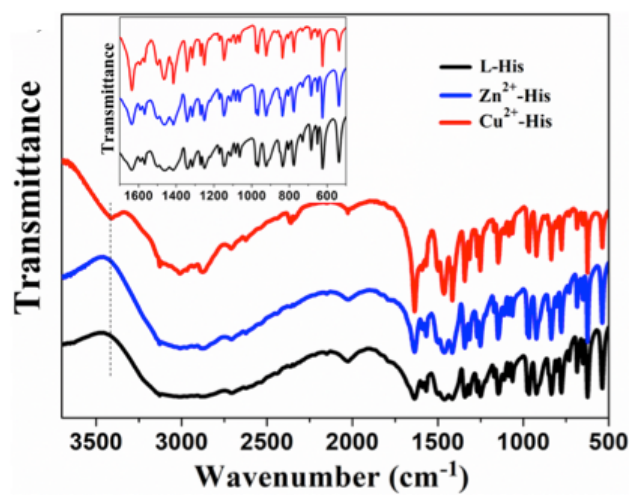

**Figure S1.** IR spectrum of $Cu^{2+}$-His solution, $Zn^{2+}$-His solution and the control group (His in pure water).

**PS2:** Fluorescence spectra of the His, $ZnCl_2$, and $Zn^{2+}$-His

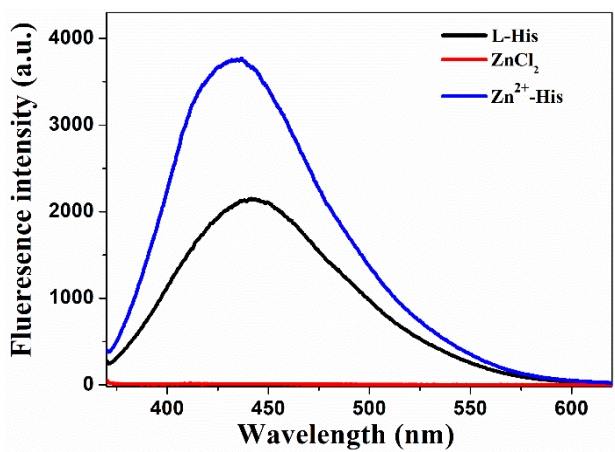

**Figure S2.** Fluorescence spectra of the His (30 mg mL$^{-1}$), $ZnCl_2$ (50 mM) in water solution and the His (30 mg mL$^{-1}$) in $ZnCl_2$ solution (50 mM).

**PS3:** UV absorption spectra of $Cu^{2+}$–Trp solution

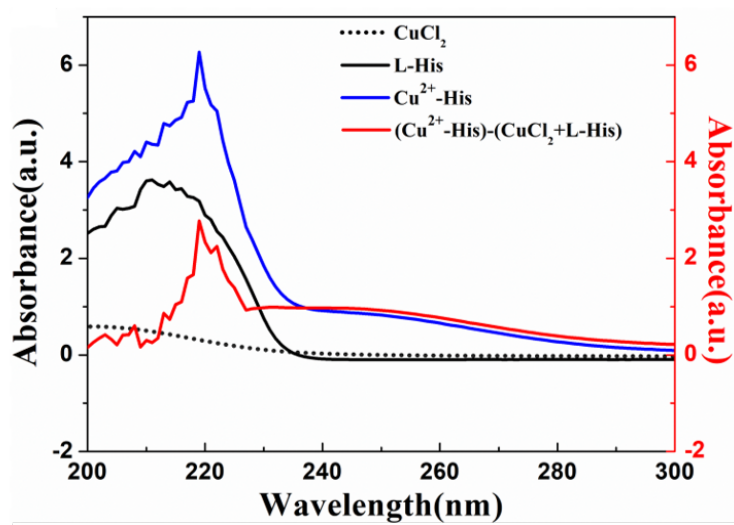

**Figure S3.** UV absorbance spectrum of $Cu^{2+}$-His solution.

## PS4: Experimental section

**Materials:**

L-Histidine (L-His, 99%), L-tryptophan (Trp, 99%), Methionine (Met, 99%), and zinc(II) chloride ($ZnCl_2$, 98%) were purchased from J&K Scientific Ltd. Copper(II) chloride dihydrate ($CuCl_2 \cdot 2H_2O$) (99%) was purchased from Sinopharm Chemical Reagent Co., Ltd. All samples were used without purification and preprocessing. All salt solutions and other aqueous solutions were prepared with 18.2 MΩ, 3 ppb TOC Milli-Q water (Millipore, US).

**Solubility measurement:**

L-His powder was added into pure water (or into the given 0.01-0.4 M $CuCl_2/ZnCl_2$ solution) with constant shaking until apparent saturation (with some insoluble L-His powder appeared), and then the solution was kept continuously stirring for 24 h in a thermostat at 25 ºC. Afterward, the solution was centrifuged at 25 ºC to remove insoluble L-His powder. Then, approximately 1 mL of solution was withdrawn by a pipette from the supernatant phase and transferred to a clean and weighed centrifuge tube. These centrifuge tubes were then transferred to liquid nitrogen for freezing and lyophilized overnight in a vacuum flask at 0.125 mbar and -50 ºC in a freeze-dryer (Virtis Freezer Dryer). The drying process was repeated until a constant mass reading was achieved. The solubility of L-His was calculated by the mass value difference of the centrifuge tube after removing salt mass. The data reported in this work was ensured by measurement of solubility for at least three replicate experiments at all compositions (Table S1 and Table S2). Based on our measurement strategy, the solubility of L-His in pure water is 41.9 mg mL$^{-1}$, consistent with previous reports[1].

**Table S1.** The pH values of pure water, 0.01 M, 0.05 M, 0.1 M, 0.2 M, 0.3 M, and 0.4

M $CuCl_2$ aqueous solution, and the solubilities of the L-His in those solvents.

| $Cu^{2+}$ concentrations (M) | Solubilities of His (mg/mL) | pH |
|---|---|---|
| 0 | 41.07±0.32 | 5.75±0.04 |
| 0.01 | 50.16±0.42 | 4.46±0.21 |
| 0.05 | 73.02±0.41 | 3.99±0.10 |
| 0.1 | 101.38±0.44 | 3.78±0.03 |
| 0.2 | 150.32±0.73 | 3.45±0.10 |
| 0.3 | 202.76±5.31 | 3.26±0.11 |
| 0.4 | 244.44±4.50 | 3.04±0.15 |

**Table S2.** The pH values of pure water, 0.01 M, 0.05 M, 0.1 M, 0.2 M, 0.3 M, and 0.4 M $ZnCl_2$ aqueous solution, and the solubilities of the L-His in those solvents.

| $Zn^{2+}$ concentrations (M) | Solubilities of His (mg/mL) | pH |
|---|---|---|
| 0 | 41.07±0.32 | 5.75±0.04 |
| 0.01 | 46.07±0.17 | 5.89±0.48 |
| 0.05 | 72.18±0.23 | 5.82±0.46 |
| 0.1 | 100.42±1.87 | 5.65±0.43 |
| 0.2 | 148.96±7.67 | 5.58±0.11 |
| 0.3 | 170.07±1.80 | 5.40±0.18 |
| 0.4 | 233.41±12.80 | 5.24±0.22 |

**Instruments:**

**Measurement of pH**

The pH was measured by SevenCompact™ S220 pH meter (pH=0~14).

**UV spectroscopy**

UV absorption spectra of His, $CuCl_2$, $ZnCl_2$, His ($CuCl_2$) and His ($ZnCl_2$) solutions were recorded on a U-3100 spectrophotometer (Hitachi, Japan). The concentration for His used in this test is 30 mg mL$^{-1}$, for $CuCl_2$ and $ZnCl_2$ is 50 mM.

### Spectrofluorophotometer

Excitation and photoluminescence (PL) spectra were measured with a Hitachi 7000 fluorescence spectrophotometer and emission slit width of 10 mm was used to record fluorescence spectra, and the fluorescence spectra of the work were recorded with $\lambda_{ex}/\lambda_{em}$= 360 nm/440 nm. The thickness of all liquid sample cells is 10 mm.

### IR spectra

Infrared (IR) spectra from 4000 cm$^{-1}$ to 500 cm$^{-1}$ of L-His, Zn$^{2+}$-His and Cu$^{2+}$-His powder were measured using a Bio-Rad FTIR spectrometer FTS165 equipped with resolution of 4 cm$^{-1}$. The drying L-His, Zn$^{2+}$-His and Cu$^{2+}$-His powder were calculated by freezing and lyophilized from corresponding saturated solution.

### Solid-State NMR Spectroscopy

Solid-state NMR experiments were carried out on a wide-bore Bruker AVANCE-600 spectrometer (14.1 T) and a DSX-400 spectrometer (Karlsruhe, Germany) on 4-mm triple-resonance MAS probes. The drying L-His, Zn$^{2+}$-His and Cu$^{2+}$-His powder were calculated by freezing and lyophilized from corresponding saturated solution.

## PS5: Computational methods

The B3LYP[2] functional in the framework of DFT is used to calculate the Cu$^{2+}$-His/Trp-H$_2$O and Zn$^{2+}$-His/Trp-H$_2$O systems. Conformation was first optimized by Berny algorithm[3] with the convergence criteria of a maximum step size of 0.0018 au and a root mean square force of 0.0003 au. A hybrid pseudo potential LanL2DZ is employed to calculate Cu$^{2+}$ and Zn$^{2+}$, while other atoms are calculated at the 6-31+G(d,p) basis set level. All calculations are done using Gaussian 09 packages[4].

Interaction between water and aromatic ring of amino acid (AA) with and without metal ion (M) are represented as $\Delta G_{abs}^{M}$ and $\Delta G_{abs}$. They can be calculated as,

$$\Delta G_{abs}^{M} = G_{Total}^{M} - G_{AA-M} - G_{w},$$

$$\Delta G_{abs} = G_{Total} - G_{AA} - G_{w}.$$

where, $G_{Total}^{M}$, $G_{Total}$, $G_{AA-M}$, $G_{AA}$ and $G_{w}$ are the single-point energies of His(Trp)-M$^{2+}$-H$_2$O, His(Trp)-H$_2$O, His(Trp)-M$^{2+}$, His and H$_2$O, respectively.